\documentclass[12pt]{article}

\usepackage{a4}
\usepackage{a4wide}

\usepackage{amsmath}
\usepackage{amscd}
\usepackage{amsfonts}
\usepackage{amssymb}
\usepackage{mathrsfs}
\usepackage{fancybox} 
\usepackage{enumerate}

\usepackage{cite}
\usepackage{bm}
\usepackage{amscd}
\usepackage{graphicx}
\usepackage[dvips]{color}
\usepackage{epsfig}

\makeatletter

\@addtoreset{equation}{section}
\makeatother

\usepackage{amsthm} 

\def\tr{\mathrm{tr}}

\def\half{{1\over2}}

\newcommand{\wt}{\widetilde}

\def\={\stackrel{\bullet}{=}}

\def\({\left(}
\def\){\right)}
\def\[{\left[}
\def\]{\right]}

\def \be {\begin{equation}}
\def \ee {\end{equation}}
\def \beqa {\begin{eqnarray}}
\def \eeqa {\end{eqnarray}}
\def \beal#1 {\begin{align}#1\end{align}}
\def \bes#1 {\begin{equation}\begin{split}#1\end{split}\end{equation}}
\def \nn {\notag\\}

\def\aver#1{\left\langle #1 \right\rangle}

\makeatletter

\@addtoreset{equation}{section}
\makeatother

\begin{document}

\begin{titlepage}
\title{
\vspace{-2cm}
\begin{flushright}
\normalsize{ 
YITP-19-095 \\ 
KUNS-2774
}
\end{flushright}
       \vspace{1.5cm}
Non-relativistic Hybrid Geometry \\ with Gravitational Gauge-Fixing Term
       \vspace{0.7cm}
}
\author{
 Sinya Aoki\thanks{saoki[at]yukawa.kyoto-u.ac.jp },
\; Janos Balog\thanks{balog.janos[at]wigner.mta.hu},\; 
Shuichi Yokoyama\thanks{shuichi.yokoyama[at]yukawa.kyoto-u.ac.jp},\; 
Kentaroh Yoshida\thanks{kyoshida[at]gauge.scphys.kyoto-u.ac.jp}
\\[25pt] 
${}^{*}{}^{\ddagger}$ {\normalsize\it Center for Gravitational Physics,} \\
{\normalsize\it Yukawa Institute for Theoretical Physics, Kyoto University,}\\
{\normalsize\it Kitashirakawa Oiwake-cho, Sakyo-Ku, Kyoto, Japan}
\\[10pt]
${}^\dagger$ {\normalsize\it Institute for Particle and Nuclear Physics,}\\ 
{\normalsize\it Wigner Research Centre for Physics,}\\ 
{\normalsize\it MTA Lend\"ulet Holographic QFT Group,}\\
{\normalsize\it 1525 Budapest 114, P.O.B.\ 49, Hungary}
\\[10pt]
${}^{\S}$ {\normalsize\it Department of Physics, Kyoto University,} \\ 
{\normalsize\it Kitashirakawa Oiwake-cho, Sakyo-ku, Kyoto, Japan}
}

\date{}

\maketitle

\thispagestyle{empty}


\begin{abstract}
\vspace{0.3cm}
\normalsize
We search a gravitational system which allows a non-relativistic hybrid geometry interpolating 
the Schr\"odinger and Lifshitz spacetimes as a solution, as a continuation of the previous work employing a flow equation. 
As such a candidate an Einstein-Maxwell-Higgs system naturally arises and we verify that this system indeed 
supports the hybrid geometry with the help of a gauge-fixing term for diffeomorphism. 
As a result, this gravitational system may be interpreted as a holographic dual 
of a general non-relativistic system at the boundary. 
\end{abstract}
\end{titlepage}

\section{Introduction} 

It is a new paradigm to realize quantum theory of gravity as a hologram of a quantum field theory 
at boundary \cite{tHooft:1993dmi,Susskind:1994vu} and has attracted much interest 
since the discovery of the AdS/CFT correspondence \cite{Maldacena:1997re}, 
which admits non-trivial tests by explicit computation \cite{Witten:1998qj,Gubser:1998bc}. 
So far, various kinds of holographic duals have been proposed. 
Among them, we are interested here in gravity duals for non-relativistic (NR) theories 
based on an NR conformal symmetry called the Schr$\ddot{\rm o}$dinger symmetry 
\cite{Son:2008ye,Balasubramanian:2008dm,Duval:2008jg} and a Lifshitz scaling symmetry 
\cite{Kachru:2008yh}. (For a nice summary, see, for example, \cite{Taylor:2008tg}.) 

\medskip

From the point of view to construct a bulk gravitational system gradually from a boundary theory 
\cite{Banks:1998dd}, it is important to specify a scale at the boundary which plays the role of 
the holographic radial direction. There are several proposals to describe how the bulk radial 
direction emerges. As one approach, some of the authors of the present letter have proposed 
that a holographic direction may be described by a flow equation 
\cite{Aoki:2015dla,Aoki:2016ohw,Aoki:2016env,Aoki:2017bru} that coarse-grains operators 
in a non-local fashion \cite{Albanese:1987ds,Narayanan:2006rf}. In comparison to 
other approaches, this flow equation method has several advantages. One of them is 
to make it possible to construct a metric of the bulk geometry directly. In particular, 
it is possible to construct an AdS geometry with a general conformally flat boundary 
\cite{Aoki:2017uce} or with a quantum-mechanically corrected bulk cosmological constant 
\cite{Aoki:2018dmc} and a holographic geometry for a general NR system \cite{Aoki:2019bfb}. 

\medskip

The resulting metric of the holographic space-time for an NR system is expressed 
as a three-parameter deformation of the $d+1$-dimensional AdS geometry \cite{Aoki:2019bfb}, 
\beal{ 
ds^2 =&\ell^2 \bigg[ - \alpha {(dx^+)^2 \over \tau^4} +  {d\tau^2 +(dx^i)^2 
+2(1 + \beta) dx^+dx^- \over \tau^2} +\gamma (dx^-)^2\bigg]\,,
\label{NRhybrid}
}
where  $\ell$ is the AdS radius, $i=1,\cdots,d-2$, and  $\alpha, \beta, \gamma$ 
are real parameters satisfying 
\beal{
\alpha \geq 0\,,  \quad \alpha  \gamma +(1+\beta )^2 > 0\,. 
\label{ParameterConstraint}
}
At a general point in the parameter space, this metric describes a $d$-dimensional 
Lifshitz space-time with the dynamical critical exponent $2$ times a straight line, 
while it enhances to the Schr\"odinger space-time in $d+1$ dimensions 
when $\beta=\gamma=0$. Therefore, this geometry would deserve to be called 
an NR {\it hybrid} geometry. 

\medskip

So far, while only the metric has been computed, a gravitational system to support this geometry  
has not been identified. In particular, the treatment of the matter sector in the context of 
the flow equation method has not been discussed. The purpose of this letter is 
to resolve these issues, and to stress that the flow equation method works well 
for the matter sector as well, not restricted to the metric. This result strongly indicates that 
the flow equation method can capture gravitation beyond geometry.

\medskip

The organization of this letter is the following. 
In section \ref{Boundary}, we review how the NR hybrid geometry is obtained 
by employing the flow equation method, and propose a candidate of the dual bulk theory 
which allows the NR hybrid geometry as a solution. In section \ref{bulkdual},  
this candidate system is verified to really support the NR hybrid geometry. 
The last section is devoted to some discussion.

\section{NR hybrid geometry from boundary} 
\label{Boundary} 

\subsection{A review of the flow equation method} 

In this section, we shall give a brief review of the results of Ref.~\cite{Aoki:2019bfb} on 
how the NR hybrid geometry \eqref{NRhybrid} can be derived starting from 
a non-relativistic conformal field theory (NRCFT) by the flow equation method. 

\medskip

Let us consider a primary scalar operator $O(\vec x,t)$ with 
a general conformal dimension $\Delta$ in a $d-1$-dimensional NRCFT. 
Note that a primary field in NRCFT is complex in relation to a U(1) symmetry group 
associated with the conservation of particle number. 

\medskip 

The two point function of primary operators is almost completely determined by an NR conformal symmetry 
as
\beal{
\aver{O(\vec x_1,t_1)O^\dagger(\vec x_2,t_2)} ={1\over (t_{12})^{\Delta}}
f\left({\vec x_{12}^2 \over 2t_{12} } \right)\,, 
\label{2ptNRCFT}
}
where $t_{12}=t_1-t_2\,, \vec x_{12}=\vec x_1-\vec x_2$\,, and $f(x)$ is a function depending on 
the given theory. The function $f$ is not necessarily an eigenfunction of the mass operator, 
which is the center in the non-relativistic conformal algebra. 
Then, by introducing an extra direction, the mass operator is realized by a differential operator 
along this direction, and the two point function is now expressed as  
\beal{
\aver{O(\vec x_1,x_1^+,x_1^{-})O^\dagger(\vec x_2,x_2^+,x_2^{-})} 
= {1\over (x_{12}^+)^{\Delta}}f\(x_{12}^-+{\vec x_{12}^2 \over 2x_{12}^+ } \)\,, 
\label{2ptNRCFTx-}
}
where $x^-$ is the extra direction and we have set $x^+=t$\,. 

\medskip 

To construct the holographic geometry, let us course-grain the conformal primary operator 
by a non-relativistic free flow equation of the form  
\be 
{\partial \over \partial \eta}\phi(x;\eta) 
= (\vec\partial^2+2\partial_- \partial_+ + 2i\bar m \partial_+) \phi(x;\eta) , 
\quad \phi(x;0) = O(x)\,, 
\label{NRflow}
\ee
where $\eta$ is a smearing variable and $\bar m$ is a real parameter of mass dimension one. 
Due to the parameter $\bar m$, the two-point function of the flowed operator $\phi$ 
is free from the contact singularity in a general parametrization. 
Thus, the flowed field can be normalized as 
\begin{eqnarray}
\sigma(x;\eta) =  \frac{\phi(x;\eta)}{\sqrt{\langle \phi(x;\eta) \phi^\dagger(x;\eta) \rangle}}\,, 
\end{eqnarray}
so that $\langle \sigma(x;\eta) \sigma^\dagger(x;\eta) \rangle=1$\,. 
Then the two-point function of normalized fields is given by 
\beal{
\aver{\sigma(x_1;\eta_1) \sigma^\dagger(x_2;\eta_2) } 
=& \left({4\eta_1\eta_2 \over \eta_+^2}\right)^{\Delta/2} 
G\left({2( x_{12}^+ + 2i\bar m \eta_+) x_{12}^{-}+ (\vec x_{12})^2\over \eta_+}\,,  
{ x_{12}^+ \over \eta_+}\right)\,,
\label{2ptgeneralFlowEq}
}
where $\eta_+=\eta_1+\eta_2$ and $G(u,v)$ is a scalar function (depending on the original $f$) 
satisfying $G(0,0)=1$\,.  

\medskip

By using the normalized field,  the so-called metric operator can be defined as
\begin{eqnarray}
\hat g_{MN}(x;\eta) \equiv \frac{1}{2} \partial_{\{M} \sigma(x;\eta) \partial_{N\}} 
\sigma^\dagger(x;\eta)\,,
\label{metricOperator}
\end{eqnarray}
where the bracket $\{~,~\}$ denotes symmetrizing the indices like 
$\{A,B\} := AB + BA$.

\medskip

The proposal is that the vacuum expectation value of the metric operator (\ref{metricOperator}) provides 
the metric of the bulk space, 
\begin{eqnarray}
ds^2=\aver{\hat g_{MN}}dz^Mdz^N\,, 
\end{eqnarray}
where $z^M$ are 
the coordinates describing the bulk space. Note that the flow variable $\eta$ is related to 
the holographic radial coordinate. The resulting metric is given by  
\beal{ 
ds^2 =& {\Delta \over 4\eta^2} d\eta^2  + {-G^{(0,2)}(\vec0) \over 
4\eta^2}(dx^+)^2 + 2 {-G^{(1,0)}(\vec0) -2i\bar m G^{(1,1)}(\vec0) \over \eta} dx^+dx^- \nn
&+(4\bar m)^2G^{(2,0)}(\vec0) (dx^-)^2 +{-\delta_{ij} G^{(1,0)}(\vec0) \over \eta} dx^idx^j\,, 
\label{generalNRgeometry}
}
where
\begin{eqnarray}
G^{(n,m)}(\vec 0) & \equiv &\left.  \frac{\partial^n}{\partial X^n} 
\frac{\partial^m}{\partial Y^m}G(X,Y)\right\vert_{X=Y=0}\,.
\end{eqnarray}
There are constraints coming from the flow equation \eqref{NRflow} like 
\begin{eqnarray}
-\Delta = 2(d-1){G}^{(1,0)}(\vec0)  + {8i\bar m }{G}^{(1,1)}(\vec0) 
+ {2i\bar m }{G}^{(0,1)}(\vec0	)\,.	
\end{eqnarray}
By performing a coordinate transformation 
\begin{eqnarray}
\eta =- {G^{(1,0)}(\vec0) \over \Delta} \tau^2\,, 
\end{eqnarray} 
the metric (\ref{generalNRgeometry}) can be rewritten to the form (\ref{NRhybrid}) 
under the identification 
\begin{eqnarray}
\alpha = {\Delta G^{(0,2)}(\vec0) \over 4G^{(1,0)}(\vec0)^2 }\,, \quad
\beta = { 2i\bar m G^{(1,1)}(\vec0) \over G^{(1,0)}(\vec0) }\,, \quad
\gamma = {4\bar m^2 G^{(2,0)}(\vec0) \over \Delta}\,,
\quad
\ell^2 = \Delta\,. 
\label{G2NRH}
\end{eqnarray}
Note that $\tau$ is the radial direction in AdS.

\subsection{Searching for the dual bulk theory} 

It is natural to ask if there exists a gravitational theory which exhibits 
the NR hybrid geometry \eqref{NRhybrid} as a solution to the equations of motion. 
To look for the dual bulk theory it is important to find out the so-called pre-geometric operators, which convert to geometric objects after taking the expectation value \cite{Aoki:2018dmc}.
If there exists a symmetry group such as $O(N)$ whose rank is taken to be large for the emergence of the bulk, these operators are constructed so as to be invariant under its global transformation.  

\medskip

In the current situation the phase rotation group plays this role, and operators invariant under the action are not only the spin-2 operator $\hat g_{\mu\nu}$ but also the following spin-1 operator 
\begin{eqnarray}
\hat A_\mu(x;\eta)\equiv {i\over2}( \partial_\mu \sigma(x;\eta) \sigma^\dagger(x;\eta) -\sigma(x;\eta) \partial_\mu \sigma^\dagger(x;\eta))\,.
\end{eqnarray}
Due to the fact that a non-relativistic conformal primary field takes values in the complex numbers, the vacuum expectation value of the spin-1 operator is non-trivial: 
\begin{eqnarray}
A_-(z) = -4\bar m G^{(1,0)}(\vec0)\,,  \qquad 
A_+(z) =- {i\Delta G^{(0,1)}(\vec0) \over 2 G^{(1,0)}(\vec0) \tau^2}\,, 
\label{spin1vev}
\end{eqnarray}
where we have set 
\begin{eqnarray}
A_\mu(z)=\langle{\hat A_\mu(x;\eta)}\rangle\,, 
\end{eqnarray}
and the other components vanish. Therefore, one can expect that not only the metric tensor $g_{\mu\nu}$ but also the gauge field $A_\mu$ may dynamically exist in the bulk theory.  

\medskip

In fact, this anticipation is consistent with a common understanding in the holography that a global symmetry 
of the boundary theory becomes a gauge symmetry of the bulk theory \cite{Witten:1998qj}.  
(See also \cite{Banks:2010zn,Harlow:2018tng}.)
To see this, let us gauge the U(1) phase rotation $O\to e^{i\lambda} O$ 
so that the parameter depends not only on the coordinates in the boundary 
but also the radial direction in the bulk. Then the spin-1 operator is transformed as 
\be 
A_\mu \to A_\mu -  \partial_\mu \lambda\,, 
\ee
which is an abelian gauge transformation. 
Thus, one can see that the spin-1 operator is converted to a gauge field in the bulk. 
This result indicates that the symmetry associated with the particle number 
corresponds to the abelian gauge symmetry in the bulk \cite{Balasubramanian:2010uw}.
Applying this argument to the flowed primary operator $\sigma$\,, 
this physical degree of freedom converts to a charged scalar field with the gauge charge unity. 

\medskip

Then a spin-2 operator defined in a suitable way should be invariant under the local U(1) transformation, while the one defined by \eqref{metricOperator} is not. There is still a possibility that extra terms can be absorbed by diffeomorphism, though our preliminary analysis suggests that it may happen by considering a special situation. 
We leave the detailed analysis of this issue to a future work.\footnote{ 
A possible way to avoid this problem is to represent the bulk metric by an information metric 
\cite{Aoki:2017bru} 
\begin{eqnarray}
ds_{\rm inf}^2 :=\frac{1}{2} \tr( d\rho(z) d\rho(z)) = g^{\rm inf}_{MN}(z) dz^M dz^N\,, 
\end{eqnarray}
where $\rho (z)$ is the density matrix defined by 
$\rho (z) := \sigma(x;\eta)\vert 0 \rangle \langle 0\vert \sigma(x;\eta)^\dagger $, and
an explicit form of $g_{MN}^{\rm inf}(z)$ can be found in the footnote 1 of \cite{Aoki:2019bfb}.
This is invariant under the local U(1) transformation preferably, whereas a metric operator whose vacuum expectation value becomes $g^{\rm inf}_{MN}(z)$ seems to be absent, unless the large $N$ factorization occurs, which causes a trouble, for instance, when correlation functions of gravitons are computed from the boundary by using this framework. 
}

\medskip

As a result, one may argue that the field contents in the bulk theory are the metric, an abelian gauge field and a charged scalar field, and thus the dual bulk theory should be a low energy effective action 
describing the dynamics of these fields.  
\begin{table}[htb]
 \label{FieldContents}
 \begin{center}
  \begin{tabular}{|cc|cc|c|}
  \hline
\multicolumn{2}{|c|}{Operators in NRCFT} &\multicolumn{2}{c|} {Bulk dynamical fields}  \\ 
  \hline
Flowed primary operator & $\sigma$  & Charged scalar field & $\Phi$ \\
Gauge field operator & $\hat A_\mu $  & Gauge field & $A_\mu$  \\
Metric operator & $\hat g_{\mu\nu} $  & Metric tensor & $g_{\mu\nu}$   \\
  \hline
  \end{tabular}
 \caption{An anticipated correspondence between the bulk dynamical fields and the boundary operators. 
 }
 \end{center}
\end{table}
A natural candidate of the system may be the Einstein-Hilbert action minimally 
coupled to an abelian gauge field as well as a charged scalar field with a certain potential. 
This system has been investigated before to study holographic duals of condensed matter systems. 
(See for instance \cite{McGreevy:2009xe,Horowitz:2010gk,Iqbal:2011ae} and references therein.)

\medskip

Indeed, one can show that the Einstein-Maxwell-Higgs system is reduced to the Einstein theory 
coupled to a massive vector field by choosing a wine-bottle potential and making the scalar field condensed. 
Then it is known that the Einstein-massive vector model admits the Schr\"odinger space-time 
\cite{Son:2008ye,Balasubramanian:2008dm,Taylor:2008tg}. 
In other words, although the NR hybrid geometry \eqref{NRhybrid} contains three parameters, 
the parameter $\alpha$ can be explained by the Einstein-Maxwell-Higgs system. 
However, from our preliminary analysis, the rest of the parameters $\beta$ and $\gamma$ cannot be explained no matter how the scalar potential is chosen. 
Then a question is how we can take account of $\beta$ and $\gamma$ without adding 
any physical degrees of freedom to the bulk, or whether there exists a bulk theory 
to support the NR hybrid geometry as a solution. 

\medskip

A clue to answer this question is in the origin of $\beta$ and $\gamma$\,. 
As seen from \eqref{G2NRH}, these parameters come from $\bar m$ contained 
in the flow equation. 
Since a choice of flow equation fixes how to perform course-graining of operators, a parameter in a flow equation is not a physical parameter in the theory. This implies that 
$\beta$ and $\gamma$ should be regarded as gauge degrees of 
freedom.\footnote{This is also suggested by the expectation value of the spin-1 field \eqref{spin1vev}.}  
For this reasoning, we repeat the same analysis by adding a gauge fixing term for the abelian gauge symmetry 
as well as that for diffeomorphism to the Einstein-Maxwell-Higgs system, and it turns out that 
the Einstein-Maxwell-Higgs theory with a gravitational gauge-fixing term admits the NR hybrid geometry 
with the general three parameters. We will verify this argument in the next section.

\section{The dual bulk theory } 
\label{bulkdual}

As argued in the previous section, let us consider an Einstein-Maxwell system minimally coupled 
to a charged scalar field with a certain potential accompanied 
with a gravitational gauge-fixing term 
\begin{eqnarray}
S &=&\int\! d^{d+1} x\, \sqrt{|g|} \[{1\over 2\kappa^2} (R - 2\Lambda) -{1\over 4} g^{\mu\nu} g^{\rho\sigma}F_{\mu\rho}F_{\nu\sigma} \right. \nonumber \\ 
&& \hspace*{3cm} \left. -g^{\mu\nu} D_\mu \Phi^* D_\nu \Phi - V(|\Phi|^2)-  {1 \over 2\xi } (g^{--})^2 \] .
\label{EMH}
\end{eqnarray}
The first term is the Einstein-Hilbert term in the Einstein frame with a negative cosmological constant 
$\Lambda$ given by%
\footnote{ 
In principle there is no way to distinguish between a cosmological constant and the expectation value of the potential physically. 
However in a situation where NRCFT is obtained from a parent CFT by a certain deformation, it would be natural to think that the cosmological constant is unchanged under the deformation and the deviation of the vacuum energy of NRCFT from that of CFT is accounted for by a newly added term, namely the value of the scalar potential. Thus we assume the same value of the cosmological constant of the AdS in \eqref{CC}.
}
\be 
\Lambda = - { d (d-1 ) \over 2\ell^2}\,. 
\label{CC}
\ee
The second term is the Maxwell term with the canonical normalization. Note that the gauge coupling constant is normalized to be unity by employing the Weyl transformation.
The third one is the kinetic term of a charged scalar field and the covariant derivative is defined as 
\be 
D_\mu \Phi \equiv \partial_\mu \Phi + i A_\mu\Phi\,, 
\ee
where the gauge charge of the scalar field was fixed as unity as expected from the boundary theory.%
\footnote{ 
In fact we cannot fix the gauge charge only by requesting the theory to enjoy a given geometry. In order to fix it, we need more information such as correlation functions or scattering amplitudes of the charged field. 
}
The fourth term is the potential of the scalar field, which is supposed to have a minimum value away from the origin so that the gauge symmetry is spontaneously broken when the scalar field condensates at the potential minimum. 
The last term is a gauge fixing term for diffeomorphism, where $\xi$ 
is a gauge-fixing parameter.\footnote{ 
In order to obtain our final result, it is important to fix diffeomorphism by breaking 
the Lorentz invariance. For example, a Fierz-Pauli-like term ${1\over 2\xi} g_{\mu\nu}g^{\mu\nu}$ 
or a gauge fixing term for the gauge symmetry such as ${1 \over 2\xi' } (A_{-})^2$ are not relevant 
to our final result. } 

\medskip

Let us show that this system supports the non-relativistic hybrid geometry \eqref{NRhybrid} 
with the following configuration of the gauge field, 
\be 
A_+ = {a_+\over \tau^2}\,, \qquad A_{-} =a_-\,, 
\label{GaugeConfig}
\ee
where $a_\pm$ are real constants, which are related to the boundary theory by 
\be 
a_+ = -{i\Delta G^{(0,1)} \over 2 G^{(1,0)}}\,,  \qquad a_{-} =-{4\bar m } {G}^{(1,0)}\,. 
\ee
To this end, it is sufficient to show that the metric \eqref{NRhybrid} 
and the gauge field \eqref{GaugeConfig} satisfy the equations of motion. 

\medskip

The Einstein equation is given by 
\be 
G_{\mu\nu} + g_{\mu\nu} \Lambda = \kappa^2 T_{\mu\nu}\,,
\label{MetricEOM}
\ee
where $G_{\mu\nu}$ is the Einstein tensor and $T_{\mu\nu}$ is the stress-energy tensor 
for the matter fields computed as 
\beal{
T_{\mu\nu} =& T^{A}_{\mu\nu} +  T^{\Phi}_{\mu\nu}  +  T^{\rm gf}_{\mu\nu}\,, \\
T^{A}_{\mu\nu} =& -{1\over 4}g_{\mu\nu} F_{\sigma\rho}F^{\sigma\rho}  
+ g^{\rho\sigma}F_{\mu\rho}F_{\nu\sigma}\,, \\
T^{\Phi}_{\mu\nu} =& g_{\mu\nu}( -D^\rho \Phi^* D_\rho \Phi - V(|\Phi|^2) ) 
+D_{\{\mu} \Phi^* D_{\nu\}} \Phi\,, \\
T^{\rm gf}_{\mu\nu} =& g_{\mu\nu} \left( -  {1 \over 2\xi } (g^{--})^2 \right) + {1 \over \xi } 
\delta_{\mu}^{-} \delta^{-}_{\nu} g^{--}\,,
}
where the curly bracket for two indices denotes the symmetrization again. 
Note that we have included the contribution coming from the gauge fixing term in the stress-energy tensor. 

\medskip

With the parametrization
\be
\Phi=r{\rm e}^{i\theta}
\ee
the scalar contribution to the Lagrangian density becomes
\be
-g^{\mu\nu}D_\mu\Phi^* D_\nu\Phi-V(\vert\Phi\vert^2)=
-g^{\mu\nu}\big\{\partial_\mu r \partial_\nu r+ r^2W_\mu W_\nu\big\}-V(r^2),
\ee
where $W_\mu=A_\mu+\partial_\mu\theta$ 
is a gauge invariant vector field.
The equation of motion for the gauge field is given by 
\be 
{1\over \sqrt{|g|}}\partial_\mu (\sqrt{|g|} F^{\mu\nu}) = -i(\Phi^* D^\nu \Phi -  D^\nu\Phi^* \Phi)=2 r^2 W^\nu
\label{GaugeEOM}
\ee
and that of the modulus field $r$
\be
\frac{1}{\sqrt{\vert g\vert}}\partial_\mu(\sqrt{\vert g\vert}\partial^\mu r)=r\big[V^\prime(r^2)+W_\mu W^\mu\big].
\label{rhoEOM}
\ee

\medskip

For our special ansatz (\ref{GaugeConfig}) we desire to look for a solution with the modulus $r$ constant, which is realized by tuning a potential to satisfy
\be 
V^\prime(r^2)= -W_\mu W^\mu.
\ee
Interestingly, this solution is different from the position of the minimum of the scalar potential, as will be seen soon.
We do not have to specify the shape of the potential for our present purposes, but we are using only the value
of the constant $r$. Note that both in the stress-energy tensor and on the right hand side of the
equation of motion (\ref{GaugeEOM}) $W_\mu$ behaves like a massive vector boson field with mass $2 r^2$.

Plugging \eqref{NRhybrid} and \eqref{GaugeConfig} 
into the left-hand side of \eqref{GaugeEOM}, it becomes 
\be 
{1\over \sqrt{|g|}}\partial_\mu (\sqrt{|g|} F^{\mu\nu})={-2a_+ [\delta^\nu_+ (d-2)\gamma\tau^2 - \delta^\nu_- d(1+\beta) ]\over 
\ell^4 (\alpha\gamma+(1+\beta)^2) }.
\ee
In the assumed parameter region \eqref{ParameterConstraint}, 
the equation of motion for the gauge field \eqref{GaugeEOM} can be solved by
\bes{
r=& \sqrt\frac{(d-2)\alpha  \gamma +d \left(\beta +1\right)^2 }{\ell^2  
\left(\alpha  \gamma +(1+\beta )^2\right)}\,,  \\
~~~  
\theta =&- \left(a_-  -\frac{2 a_+ \gamma  (1 + \beta ) }{(d-2)\alpha  \gamma +d
\left(\beta +1\right)^2}\right) x^-\,.
\label{ScalarConfig}
}
Note that the value of $W_\mu W^\mu$ becomes constant for this solution.

\medskip

For these configurations, the stress energy tensors are computed as 
\beal{
T^{\Phi}_{\mu\nu} 
=& g_{\mu\nu}\left( \frac{(a_+)^2 \gamma  \left( (d-2)^2\alpha  \gamma
+ d(d-4)(\beta +1)^2 \right)}{\ell^4 \left(\alpha  \gamma +(\beta +1)^2\right)\left\{(d-2)\alpha\gamma + d(\beta+1)^2\right\}}  - V(r^2) \right) \nn
&+2r^2\bigg( {a_+^2 \over \tau^4} \delta_\mu^+\delta_\nu^+ + {2 \over  \tau^2}\frac{a_+^2 
\gamma (\beta +1) }{(d-2)\alpha  \gamma +d \left(\beta +1\right)^2}\delta_{\{\mu}^- 
\delta_{\nu\}}^+ \nn
&+ \left( \frac{a_+ \gamma 2(\beta +1) }{(d-2)\alpha  \gamma +d 
\left(\beta +1\right)^2}\right)^2\delta_\nu^-\delta_\mu^-\bigg)\,, \\
T^{A}_{\mu\nu}
=& \left({1\over 2}g_{\mu\nu} {\gamma \over \ell^4 \left(\alpha \gamma+ (1+\beta)^2\right)} 
+\delta_{\mu}^+\delta_{\nu}^+  {1\over \ell^2 \tau^4}+\delta_{\mu}^\tau\delta_{\nu}^\tau {-\gamma 
\over \tau^2\ell^2 \left(\alpha \gamma+ (1+\beta)^2\right)}\right) ({2a_+})^2\,, \\
T^{\rm gf}_{\mu\nu} =& g_{\mu\nu}\left( -  {1 \over 2\xi } ({\alpha \over \ell^2\left(\alpha \gamma 
+ (1+\beta)^2\right)})^2 \right) + {1 \over \xi } \delta_{\mu}^{-} \delta^{-}_{\nu} {\alpha \over 
\ell^2\left(\alpha \gamma+ (1+\beta)^2\right)}\,.
}
On the other hand, the Einstein tensor with the contribution of the cosmological constant for the metric 
\eqref{NRhybrid} is computed as 
\beal{
G_{\mu\nu} + g_{\mu\nu} \Lambda =& {\alpha \gamma \over  \alpha \gamma+ (1+\beta)^2}{g_{\mu\nu} \over \ell^2}  +\delta_{\mu}^-\delta_{\nu}^- {\gamma ((d-2)(1+\beta)^2 + d\alpha\gamma)  \over \alpha\gamma+ (1+\beta)^2} + {\delta_{\mu}^{\{+}\delta_{\nu}^{-\}} \over \tau^2} { 2(1+\beta) \alpha\gamma \over \alpha\gamma+ (1+\beta)^2  }\nn
&+{\delta_{\mu}^+\delta_{\nu}^+ \over \tau^4} { \alpha ((d+2)(1+\beta)^2 + d \alpha\gamma) \over \alpha\gamma+ (1+\beta)^2 }
+{\delta_{\mu}^\tau\delta_{\nu}^\tau \over \tau^2} {-2\alpha\gamma  \over  \alpha \gamma+ (1+\beta)^2}\,. 
}
Plugging these into the Einstein equation, we obtain five equations to be satisfied from the components $(\mu,\nu)=(i,j),~(+,+),~(+,-),~(-,-),~(\tau,\tau)$\,.
The equation for the  $(+,-)$ component is satisfied 
if 
\be 
a_+^2 = {\ell^2 \alpha \over 2\kappa^2}\,, 
\label{a+}
\ee
thus the gauge flux does not vanish, as expected.
Then the equations for the $(+,+)$ and $(\tau,\tau)$ components are automatically satisfied.
The equation for the $(-,-)$ component fixes the parameter in the gauge-fixing term as 
\beal{
{1\over\xi} =&{ d(d-2)[\alpha\gamma + (1+\beta)^2]^2 \over  (d-2)\alpha  \gamma +d \left(1+\beta \right)^2 } {\gamma\ell^2 \over \alpha \kappa^2 }\,.
\label{gaugeparam}
}
The equation for the $(i,j)$ component fixes the potential at $r^2$ as 
\beal{
V(r^2)
=- \frac{2a_+^2 \gamma}{\Delta^2(\alpha\gamma +(1+\beta)^2) }.
\label{potentialMinimum}
}
This completes the verification of our claim. 
Note that from \eqref{a+} the flow field determines the Newton constant $\kappa^2$ as  
\begin{eqnarray}
\kappa^2 = \frac{G^{(0,2)}(\vec0)}{2 (iG^{(0,1)}(\vec0))^2 }\,.
\end{eqnarray}

\medskip

It is worth commenting on the Higgs mechanism to reduce the Einstein-Maxwell-Higgs theory 
to the Einstein theory with a massive vector field. To this end, the matter field is 
expanded around the vacuum expectation value like 
\be 
\Phi = {\rm e}^{i\theta}(r + x+iy)\,,  \qquad  A_\mu = \aver{A_\mu} + \delta A_\mu\,. 
\ee
Here $r,\theta$ are given in \eqref{ScalarConfig} and $\aver{A_\mu}$ follows from 
\eqref{GaugeConfig}\,. Then $x,y, \delta A_{\mu}$ are fluctuation fields taking real values. 
Then the matter part of the action \eqref{EMH} can be expanded,    
up to the quadratic order.\footnote{ For this expansion a background dependent Lorenz gauge 
$g^{\mu\nu} (\nabla_\mu - ie\aver{A_\mu} ) \delta A_\nu = 0$ may be useful.}  
From the expansion of the potential term, we find that the field $y$ is  a massless mode 
corresponding to the Nambu-Goldstone boson and $x$ is a massive mode with 
the mass squared $4V''(r^2) r^2$\,. 
By making the potential so steep that the mass is much heavier than the energy scale 
of our interest, the dynamics of the field $x$ becomes negligible, 
while the massless mode is absorbed by the gauge field so as to become a massive vector field. 
As a result, by setting 
\be 
\wt A_\mu = \delta A_\mu + {1 \over r}(\partial_\mu + i\aver{A_\mu} +i\partial_\mu\theta )y\,, 
\qquad  
m^2 =2 r^2 , 
\ee
we obtain 
\beal{ 
S =&\int\! d^{d+1} x\, \sqrt{|g|} \[{1\over 2\kappa^2} (R - 2\Lambda) 
-{1\over 4} g^{\mu\nu} g^{\rho\sigma}\wt F_{\mu\rho}\wt F_{\nu\sigma} 
- \half m^2 g^{\mu\nu} \wt A_\mu \wt A_\nu 
-  {1 \over 2\xi } (g^{--})^2 \]\,, 
\label{MassiveVector}
}
where $\wt F_{\mu\nu} = \partial_\mu \wt A_\nu -  \partial_\nu \wt A_\mu$ and 
the irrelevant terms have been ignored. As a result, the Einstein-Maxwell-Higgs theory \eqref{EMH} 
reduces to the Einstein theory with a massive vector field \eqref{MassiveVector} 
by choosing the potential appropriately. Thus, it has been shown that the Einstein-massive vector 
theory with a suitable gauge-fixing term for diffeomorphism also supports the NR hybrid geometry \eqref{NRhybrid} 
as a solution.

\section{Discussion} 
\label{Discussion} 

In this letter we have considered the dual gravitational theory of a general NRCFT and proposed 
that the theory is the Einstein theory minimally coupled to an abelian gauge field 
and a Higgs field with a gravitational gauge fixing term, by confirming explicitly 
that the theory admits the NR hybrid geometry. 

\medskip 

In spite of our explicit verification, we have not fully understood the role of the gravitational 
gauge-fixing term in the context of holography.\footnote{We would like to thank Shigeki Sugimoto for his valuable comment on this point.}
It seems that only zero or nonzero are meaningful 
about the parameter of the gauge fixing term and induce an important effect to solve 
the equations of motion. This result may indicate that a particular choice of smearing  
would correspond to a particular gauge choice in the bulk. 
It is significant to elaborate this point in detail.

\medskip

It is also interesting to test our proposal by computing correlation functions for both sides 
and comparing them based on the dictionary in Tab.~\ref{FieldContents}.
An advantage of the flow equation method is that one can perform explicit computations  
by using traditional techniques of quantum field theory such as the $1/N$ expansion, which enables one to verify his/her proposal in an  analytic way. 

\medskip

We hope to come back to these issues in the near future.

\section*{Acknowledgment}

The authors would like to thank Shigeki Sugimoto for his valuable comment and discussion during the YITP-W-19-10 on "Strings and Fields 2019". 
S. A. is supported in part by the Grant-in-Aid of the Japanese Ministry of Education, Sciences and Technology, Sports and Culture (MEXT) for Scientific Research (No. JP16H03978),  
by a priority issue (Elucidation of the fundamental laws and evolution of the universe) to be tackled by using Post ``K" Computer, 
and by Joint Institute for Computational Fundamental Science (JICFuS).
S. Y. is supported in part by the Grant-in-Aid of the Japanese Ministry of Education, Sciences and Technology, Sports and Culture (MEXT) for Scientific Research (No. JP19K03847). 
The work of K.Y. was supported by the Supporting Program for Interaction-based Initiative 
Team Studies (SPIRITS) from Kyoto University, a JSPS Grant-in-Aid for Scientific Research (B) 
No.\,18H01214. This work is also supported in part by the JSPS Japan-Russia Research Cooperative Program.
J. B. was partially supported by the Hungarian National
Science Fund OTKA (under K116505).

\bibliographystyle{utphys}
\bibliography{NRhybrid}

\end{document}